\documentstyle[eqsecnum,aps]{revtex}

\begin{document}
\draft
\title{Massive torsion modes, chiral gravity,\\
and the Adler-Bell-Jackiw anomaly }
\author{Lay Nam Chang\cite{byline1}}
\address{
Department of Physics, Virginia Tech.\\
Blacksburg, VA24061-0435, USA. }
\author{Chopin Soo\cite{byline2}\\
}
\address{Department of Physics,
National Cheng Kung University\\
Tainan 70101, Taiwan. }

\maketitle
\begin{abstract}
Regularization of quantum field theories introduces a mass scale
which breaks axial rotational and scaling invariances. We
demonstrate from first principles that axial torsion and torsion
trace modes have non-transverse vacuum polarization tensors, and
become massive as a result.
The underlying reasons are similar to those responsible for the
Adler-Bell-Jackiw (ABJ) and scaling anomalies.
Since these are the only torsion components that can couple minimally to spin $\frac{1}{2}$ particles, the
anomalous generation of masses for these modes, naturally of the order of the regulator scale, may help to
explain why torsion and its associated effects, including CPT violation in chiral gravity, have so far escaped
detection. As a simpler manifestation of the reasons underpinning the ABJ anomaly than triangle diagrams, the
vacuum polarization demonstration is also pedagogically useful. In addition it is shown that the teleparallel
limit of a Weyl fermion theory coupled only to the left-handed spin connection leads to a counter term which is
the Samuel-Jacobson-Smolin action of chiral gravity in four dimensions.

\end{abstract}
\pacs{PACS numbers: 11.15.-q, 11.40.Ha, 04.62.+v, (Journal-ref: {Class. Quantum Grav. {\bf 20} (2003) 1379-1387)}}

\widetext
\section{Introduction and Overview}
\label{sec:level1}

   Torsion $(T_A = de_A + A_{AB}\wedge e^B$) arises naturally in
Riemann-Cartan spacetimes when the vierbein, $e_{\mu A}$, and
spin connection, $A_{\mu AB}$, are assumed to be independent. In
minimal coupling schemes only spinors couple to torsion, and even
then only the axial and trace modes of torsion couple to spin
$\frac{1}{2}$ particles. Actually, in a hermitian theory, only
the axial torsion mode, ${\tilde A}_\mu \equiv
\frac{1}{2}g_{\mu\lambda}{\tilde\epsilon}^{\lambda\nu\alpha\beta}
e_{\lambda A}T^A_{\alpha\beta}$, interacts minimally with spin
$\frac{1}{2}$ matter. Recent works have however revealed that even
the well-studied Adler-Bell-Jackiw (ABJ) anomaly \cite{ABJ}
receives further contributions from torsional invariants
\cite{Yajima,Chandia,cpabj}. Moreover, as we shall demonstrate,
the vacuum polarization diagram with two external axial torsion
vertices is not transverse, and its divergence is controlled by
the Nieh-Yan term \cite{Nieh-Yan}. This breakdown in
transversality occurs in addition to that manifested by the ABJ
triangle diagrams that give rise to a term quadratic in the
curvatures. A striking consequence of this non-transversality is
the generation of mass. We may think of the axial torsion mode as
an ``axial torsion photon" \cite{cpabj}, coupled to a current
whose conservation has been compromised because of anomaly
considerations.  The associated ``gauge" invariance, the
$\gamma^5$ rotational symmetry,  is therefore broken, and a mass
results. In this paper, we describe explicitly how this
phenomenon takes place. It should be noted that the breakdown in
current conservation poses no consistency problems; since axial
torsion modes are not gauge field modes, and are not responsible
for any local symmetries.

The standard model incorporates maximal parity and charge conjugation non-conservation by assigning left- and
right- handed fermions to different representations of the internal gauge group. It behooves us to pursue the
extent to which chirality can be used as a defining characteristic of particle interactions, including gravity.
That context leads us to use only Weyl spinors and the Ashtekar formulation of gravity \cite{Ash,cps}. The result
in using only left-handed fermion fields is a minimal coupling $J^\mu_L (i B_\mu + A_\mu) \equiv J^\mu_L C_\mu$
of both the axial torsion and torsion trace fields to the total singlet current $J^\mu_L$ traced over all
left-handed fermion fields. Here, ${A}_\mu \equiv \frac{1}{4e}{\tilde A}_\mu$, while $B_\mu \equiv
\frac{1}{2}e^{\nu A}T_{\mu\nu A}$ is the trace of the torsion field. The first term in $C_\mu$ is anti-hermitian
relative to the second, and hermitizing the action would eliminate it completely. But doing so requires bringing
in right-handed conjugate Weyl fields, which in turn are coupled to the self-dual or right-handed spin connection
fields. These fields are distinct components of the spin connection, and it has already been shown in Ref.
\cite{Ash} that general relativity field equations can be reproduced without reference to them. The system can be
defined by a holomorphic action that is dependent only upon left-handed fields, thereby extending this
characteristic feature of the standard model to cover gravity interactions as well\cite{cps,cpt}. In this scheme
of things, the appearance of $iB_\mu$ is entirely natural. In the regularization scheme adopted in this
paper\cite{invariant}, this property is also maintained, therefore both torsional modes must have the same mass.
The generation of mass for the $B$ field is not self-evident, since it has a vector coupling to the fermion
fields rather than an axial vector coupling.  Based upon our experience with Quantumelectrodynamics (QED), we
might have argued that the associated quanta remain massless. The difference arises because of the anti-hermitian
coupling. The operator appearing in the regularized integrals involves the positive definite form in Euclidean
space $i{D\kern-0.15em\raise0.17ex\llap{/}\kern0.15em\relax}
(i{D\kern-0.15em\raise0.17ex\llap{/}\kern0.15em\relax})^\dagger$. Now
$(i{D\kern-0.15em\raise0.17ex\llap{/}\kern0.15em\relax})^\dagger$ differs from
$i{D\kern-0.15em\raise0.17ex\llap{/}\kern0.15em\relax}$, because of the anti-hermitian nature of the coupling.
For conventional internal gauge fields, the two forms of the covariant derivatives are the same, leading to
hermitian Dirac operators. We shall show how this difference causes the vacuum polarization tensor for $B_\mu$ to
be non-transverse as well.


\subsection{Vacuum polarization and the ABJ Current}

Consider first the action of a bispinor theory in teleparallel
spacetimes with flat vierbein $e^A_{\mu} = \delta^A_\mu$ but with
nontrivial axial torsion coupling
\begin{equation}
S = - \frac{1}{2}\int d^4x \, e{\overline \Psi}\gamma^\mu
(i\partial_\mu + {A}_\mu\gamma^5 )\Psi + H. c. \label{flataction}
\end{equation}
The ABJ current $J^{5\mu} ={\overline\Psi}\gamma^\mu \gamma^5\Psi$ has naive expectation value
\begin{equation}
\langle J^{5\mu}(x) \rangle = -\lim_{x \rightarrow y} {\rm
Tr}\{\gamma^\mu \gamma^5 {1 \over
{(i{\partial\kern-0.15em\raise0.17ex\llap{/}\kern0.15em\relax} +
{A\kern-0.15em\raise0.17ex\llap{/}\kern0.15em\relax}\gamma^5)}}\delta(x-y)\},
\end{equation}
and the corresponding vacuum polarization tensor, $\Pi^{\mu \nu}$,
defined by the Fourier transform
\begin{equation}
\left.{{\delta \langle J^{5\mu}(x) \rangle}\over {\delta
A_\nu(y)}}\right|_{A_{\alpha}=0} = \int {{d^4k} \over {(2\pi)^4}}
\Pi^{\mu\nu}(k) e^{ik.(x-y)}, \label{e3}
\end{equation}
leads to
\begin{equation}
\Pi^{\mu\nu}(k) \propto \int d^4p \, {\rm Tr}\{\gamma^\mu\gamma^5
{1 \over {{p\kern-0.15em\raise0.17ex\llap{/}\kern0.15em\relax} +
{k\kern-0.15em\raise0.17ex\llap{/}\kern0.15em\relax}}}
\gamma^\nu\gamma^5 {1\over
{{p\kern-0.15em\raise0.17ex\llap{/}\kern0.15em\relax}}}\}.
\end{equation}
Had this expression been well defined, it would have been no more
than
\begin{equation}
\Pi^{\mu\nu}(k) \propto \int d^4p  \, {\rm Tr}\{\gamma^\mu {1
\over {{p\kern-0.15em\raise0.17ex\llap{/}\kern0.15em\relax} +
{k\kern-0.15em\raise0.17ex\llap{/}\kern0.15em\relax}}} \gamma^\nu
{1\over {{p\kern-0.15em\raise0.17ex\llap{/}\kern0.15em\relax}}}\}
\end{equation}
since the two $\gamma^5$'s cancel out in the trace, and we would
have obtained the usual vacuum polarization amplitude for which
we do not expect a longitudinal component. But it is not, for the
integration over fermion loop momentum diverges. We will need a
regularization scheme, Pauli-Villars for instance, to tame this
divergence before performing any Dirac algebra. Any
gauge-invariant scheme however will compromise symmetry generated
by the axial current. Summing over the propagators for all the
fields, including the massive regulators, results in
\begin{equation}
\Pi^{\mu\nu}(k) \propto \sum_n C_n \int d^4p \, {\rm
Tr}\{\gamma^\mu\gamma^5 {1 \over
{({p\kern-0.15em\raise0.17ex\llap{/}\kern0.15em\relax} +
{k\kern-0.15em\raise0.17ex\llap{/}\kern0.15em\relax}) + i m_n }}
\gamma^\nu \gamma^5 {1 \over
{{p\kern-0.15em\raise0.17ex\llap{/}\kern0.15em\relax} + i m_n}}\},
\end{equation}
with $C_n = \pm 1$, depending on whether the regulators are
anti-commuting or commuting. (For the original fermion multiplet,
$C_0 =1$ and $m_0= 0$. We assume analytic continuation to
Euclidean Green functions.) By moving the second $\gamma^5$ to the
left to cancel out the first, we observe that $m_n$ {\it changes
its relative sign} with respect to
$({p\kern-0.15em\raise0.17ex\llap{/}\kern0.15em\relax} +
{k\kern-0.15em\raise0.17ex\llap{/}\kern0.15em\relax})$ in the
denominator. Consequently,
\begin{equation}
\Pi^{\mu\nu}(k) \propto \sum_n C_n \int d^4p \, {\rm
Tr}\{\gamma^\mu{1 \over
{({p\kern-0.15em\raise0.17ex\llap{/}\kern0.15em\relax} +
{k\kern-0.15em\raise0.17ex\llap{/}\kern0.15em\relax}) -i m_n }}
\gamma^\nu {1 \over
{{p\kern-0.15em\raise0.17ex\llap{/}\kern0.15em\relax} +i m_n}}\}.
\label{e1}
\end{equation}
The integrals over the loop momentum will be well defined for a
suitable set $\{C_n, m_n\}$. An explicit set will be presented
for the Weyl theory. It is also applicable to the bispinor theory
here, consistently yielding a polarization magnitude which is
twice that of the Weyl theory. The important point is that if we
had started with a vector (instead of the axial vector) coupling,
the result for Eq.\ (\ref{e1}) would have been
\begin{equation}
\Pi^{\mu\nu}(k) \propto \sum_n C_n \int d^4p \, {\rm
Tr}\{\gamma^\mu {1 \over
{({p\kern-0.15em\raise0.17ex\llap{/}\kern0.15em\relax} +
{k\kern-0.15em\raise0.17ex\llap{/}\kern0.15em\relax}) +i m_n }}
\gamma^\nu {1 \over {{p\kern-0.15em\raise0.17ex\llap{/}
\kern0.15em\relax} +i m_n}}\} \label{e2}
\end{equation}
instead.
This integral would have produced a transverse polarization
tensor.  As it is, by rewriting one of the propagators in Eq.\
(\ref{e1}) as
\begin{equation}
{1 \over {({p\kern-0.15em\raise0.17ex\llap{/}\kern0.15em\relax} +
{k\kern-0.15em\raise0.17ex\llap{/}\kern0.15em\relax}) -i m_n}} =
{1 \over {({p\kern-0.15em\raise0.17ex\llap{/}\kern0.15em\relax} +
{k\kern-0.15em\raise0.17ex\llap{/}\kern0.15em\relax}) +i m_n}} -
{2i m_n \over {(p + k)^2 + m^2_n}}.\label{two}
\end{equation}
we obtain two terms, the first of which is identical to the
integral in Eq.\ (\ref{e2}).  However,
there is now an additional anomalous term
\begin{equation}
\sum_n C_n \int d^4p \, {\rm Tr}\{\gamma^\mu {2i m_n \over {(p +
k)^2 + m^2_n }} \gamma^\nu
{({p\kern-0.15em\raise0.17ex\llap{/}\kern0.15em\relax} -i m_n)
\over {p^2 + m^2_n}}\} = 8g^{\mu\nu} \sum_n C_n \int d^4p
\,{{m^2_n} \over {[(p + k)^2 + m^2_n][p^2 + m^2_n] }}.
\end{equation}
Clearly the anomalous component arises from both the axial vector
coupling and the nontrivial regulator masses. From
\begin{equation}
\Pi^{\mu\nu} = (k^\mu k^\nu - g^{\mu\nu} k^2)\Pi + g^{\mu\nu}\Pi',
\end{equation}
we deduce
\begin{equation}
k_\mu \Pi^{\mu\nu} \propto k^\nu \quad {\rm and} \quad \langle
\partial_\mu J^{5\mu}\rangle_{\rm Reg} \propto \partial_\mu
{\tilde A}^\mu \neq 0
\end{equation}
at the level of vacuum polarization diagrams. The anomalous
$g^{\mu\nu}\Pi'$ contribution in the vacuum polarization tensor
also implies that besides the usual $F_{\mu\nu}F^{\mu\nu}$ piece
required for the logarithmic divergence of transverse
polarization, a mass counter term of the form $A_\mu A^\mu$ is
also needed for the non-zero longitudinal component in the
effective action. $A_\mu$ becomes massive as a result.

In vector QED, curbing divergences by naive momentum truncation
also results in a non-transverse photon polarization tensor
\cite{PeskinBook}. But this apparent breakdown of gauge
invariance is an artifact of symmetry breaking ``regularization"
which can be removed altogether by proper gauge-invariant
regularization schemes. In contradistinction, the non-transverse
torsion polarization exhibited here is not the consequence of a
``fake anomaly" resulting from ``improper regularization" which
breaks the symmetry of axial rotations. The complete ABJ anomaly
assures us there are no regularization schemes that preserve
singlet axial rotations, and at the same time respect all of the
local symmetries, such as Lorentz and other gauge symmetries,
which are present. The non-transverse polarization for axial
torsion can thus be regarded as another manifestation of this
phenomenon. $A_\mu$ however transforms covariantly under
diffeomorphisms, and is Lorentz invariant. The counter term
necessary to compensate for a non-transverse polarization tensor,
$A_\mu A^\mu$, is therefore completely consistent with these local
invariances. A similar term in QED would have violated local gauge
invariance.

It is also instructive to exhibit and confirm the same effect
using an alternative regularization method. For bispinors, we may
write the effective action in curved space as
\begin{eqnarray}
\Gamma_{\rm eff.} &=& -i {\rm Tr}\ln [e^{1\over 2}i
{\Delta\kern-0.15em\raise0.17ex\llap{/}\kern0.15em\relax}
e^{-{1\over 2 }}], \cr
\nonumber\\
i{\Delta\kern-0.15em\raise0.17ex\llap{/}\kern0.15em\relax} &=&
\gamma^\mu(i\partial_\mu + {i\over 2}\omega_{\mu AB}\sigma^{AB} +
A_\mu \gamma^5).
\end{eqnarray}
With heat kernel regularization and the Schwinger-DeWitt
expansion\cite{DeWitt}, the ABJ anomaly with Euclidean signature
has been demonstrated to take the form\cite{Yajima}
\begin{eqnarray}
\langle \partial_\mu J^{5\mu}\rangle_{\rm Reg.} &=& 2i\lim_{t
\rightarrow 0}\lim_{x \rightarrow x'}{1\over{(4\pi t)}^2} {\rm
Tr}\langle x'|\gamma^5
\exp[-t(i{\Delta\kern-0.15em\raise0.17ex\llap{/}\kern0.15em\relax})^2]
|x\rangle \cr
\nonumber\\
&=& 2i\lim_{t \rightarrow 0}{1\over{(4\pi t)}^2} {\rm
Tr}[\sum^\infty_{n=0} e\gamma^5 a_n t^n].
\end{eqnarray}
Furthermore, it is known that the traces of the coefficients
$a_0, a_1$ and $a_2$ contribute to the divergent part of the
effective action. For instance, $a_0 = I$ leads to the
renormalization of the cosmological constant. We focus on the
relevant coefficient $a_1$ which, in our notation,
is $a_1 = -\frac{1}{12}R -2A_\mu A^\mu - \gamma^5
e^{-1}\partial_\mu(eA^\mu)$. Clearly ${\rm Tr}(e a_1)$ (and thus
the effective action) contains both the familiar Einstein-Hilbert
term, $e R$, as well as the $e A_\mu A^\mu$ mass term which we
also found to be required by the vacuum polarization computations
above. It follows from Eq.(1.14) that the ABJ anomaly is
\begin{equation}
\langle \partial_\mu J^{5\mu}\rangle_{\rm Reg.} = -{{2i}\over
{(4\pi)^2 t}}\partial_\mu \tilde A^\mu + {{2i}\over
{(4\pi)^2}}{\rm Tr}(e\gamma^5 a_2).
\end{equation}
${\rm Tr}(e\gamma^5 a_2)$ is the more familiar regulator scale
independent part of the ABJ anomaly. But there is also the
$t$-dependent first term ($t$ has the physical dimension of
inverse regulator mass squared). This is precisely the Nieh-Yan
contribution to the ABJ anomaly.
The linear dependence on $A_\mu$ shows that
the vacuum polarization is indeed the correct Feynman diagram
process to consider for this purpose\cite{cpabj}. Therefore the
non-transversality of the polarization tensor is not only
genuine, but is in fact necessary to understand the origin of the
Nieh-Yan contribution to the ABJ anomaly in perturbation theory.

For the truly chiral Weyl theory which we shall focus on next, the
``chiral determinant" is actually a section of the determinant
line bundle, and the effective action is not a straightforward
functional determinant. Nevertheless, it is feasible to compute
the polarization of torsion fields while preserving explicit
left-handedness as well as Lorentz and gauge invariance.
Chirality of the gravitational couplings and the ABJ anomaly
ensure the combination $C_\mu = iB_\mu + A_\mu$ exhibits
non-transverse polarization.


\section{Weyl fermions and Vacuum Polarization tensor of torsion fields}

The classical action for a left-handed multiplet
${\overline\Psi}_L$ consisting of $d$ Weyl fermions is
\begin{equation}
S_L = -\int d^4x\, e{\overline\Psi}_L
i{D\kern-0.15em\raise0.17ex\llap{/}\kern0.15em\relax}\Psi_L,
\end{equation}
with $i{D\kern-0.15em\raise0.17ex\llap{/}\kern0.15em\relax} =
\gamma^\mu(i\partial_\mu + \frac{i}{2}A_{\mu AB}\sigma^{AB} +
W_{\mu a}{T}^{a})$. It is known that if the representation of the
internal gauge field $T^a$ is perturbatively anomaly-free $({\rm
Tr}(T^a) ={\rm Tr}(T^a\{T^b,T^c\}) = 0)$, then all fermionic loops
in background gauge and gravitational fields of the theory can be
regularized in an explicitly gauge, Lorentz and diffeomorphism
invariant manner. To explicitly maintain the chiral gravitational
couplings we utilize an infinite tower of Pauli-Villars
regulators doubled in the internal space (please see Ref.
\cite{invariant} for further details.) This generalizes the
invariant scheme first introduced by Frolov and
Slavnov\cite{Slavnov}. Specifically, to form invariant regulator
masses, the internal space is doubled from $T^a$ to
\begin{equation}
{\cal T}^a =\left(\matrix{(-T^a)^* &0\cr 0&T^a}\right);
\end{equation}
and the original fermion multiplet is projected as $\Psi_{L} =
\frac{1}{2}(1-\sigma^3)\Psi_{L}$, where
\begin{equation}
\sigma^3 =\left(\matrix{1_d &0\cr 0& -1_d}\right).
\end{equation}
The chirality of the regularized theory with respect to the
gravitational interaction is thus preserved even in the
teleparallel limit. It can be shown\cite{invariant} the net effect
of the regularization is to replace the $\frac{1}{2}(1-\sigma^3)$
projection of the bare currents by
$\frac{1}{2}(f({{{D\kern-0.15em\raise0.17ex\llap{/}\kern0.15em\relax}
{{D\kern-0.15em\raise0.17ex\llap{/}\kern0.15em\relax}^\dagger}}
\over{\Lambda^2}})-\sigma^3)$ where $f$ is the regulator function,
\begin{equation}
f({D\kern-0.15em\raise0.17ex\llap{/}\kern0.15em\relax}
{{D\kern-0.15em\raise0.17ex\llap{/}\kern0.15em\relax}}^\dagger/\Lambda^2)
=\sum_n C_n{{i{D\kern-0.15em\raise0.17ex\llap{/}\kern0.15em\relax}
({i{D\kern-0.15em\raise0.17ex\llap{/}\kern0.15em\relax}})^\dagger}
\over{[i{D\kern-0.15em\raise0.17ex\llap{/}\kern0.15em\relax}
{(i{D\kern-0.15em\raise0.17ex\llap{/}\kern0.15em\relax})^\dagger}
+ m^2_n]}} =\sum^{\infty}_{n= -\infty} {{(-1)^n
i{D\kern-0.15em\raise0.17ex\llap{/}\kern0.15em\relax}
({i{D\kern-0.15em\raise0.17ex\llap{/}\kern0.15em\relax}})^\dagger}
\over{[i{D\kern-0.15em\raise0.17ex\llap{/}\kern0.15em\relax}
{(i{D\kern-0.15em\raise0.17ex\llap{/}\kern0.15em\relax})^\dagger}
+ n^2\Lambda^2]}}.
\end{equation}

To concentrate on the vacuum polarization of the torsion fields,
we specialize to $W_{\mu a} =0$, and flat vierbein $e_{\mu A}
=\eta_{\mu A}$ but retain nontrivial torsion couplings in the
teleparallel limit. To wit, the Weyl action reduces to
\begin{equation}
S_L= \int d^4x [-e{\overline\Psi}_L\gamma^\mu i\partial_\mu\Psi_L
+ C_\mu{\overline\Psi}_L e\gamma^\mu \Psi_L]. \label{Laction}
\end{equation}
The bare current
\begin{equation}
\langle {{\delta S_L}\over{\delta C_{\nu}(x)}}\rangle =\langle
J^\mu_L \rangle_{Bare} =\lim_{x \rightarrow y} {\rm
Tr}\{\gamma^\mu\frac{1}{2}(1-\gamma^5)
{1\over{i{D\kern-0.15em\raise0.17ex\llap{/}\kern0.15em\relax}}}
\frac{1}{2}(1-\sigma^3)\delta(x-y)\}
\end{equation}
is modified by the Pauli-Villars regulators to become
\begin{eqnarray}
\langle J^{\mu}_L(x)\rangle_{\rm Reg} &=& \lim_{x \rightarrow
y}{\rm Tr}\{\gamma^\mu \frac{1}{2}(1-\gamma^5) {1\over
{i{D\kern-0.15em\raise0.17ex\llap{/}\kern0.15em\relax}}}
\frac{1}{2}(f -\sigma^3)\delta(x-y)\}\cr
\nonumber\\
&=&\lim_{x \rightarrow y}{\rm
Tr}\{\gamma^\mu\frac{1}{2}(1-\gamma^5) (\{\frac{1}{2}\sum_n
{{(-1)^n
({i{D\kern-0.15em\raise0.17ex\llap{/}\kern0.15em\relax}})^\dagger}
\over{[i{D\kern-0.15em\raise0.17ex\llap{/}\kern0.15em\relax}
{(i{D\kern-0.15em\raise0.17ex\llap{/}\kern0.15em\relax})^\dagger}
+ m^2_n]}}\}
-{1\over{2i{D\kern-0.15em\raise0.17ex\llap{/}\kern0.15em\relax}}}\sigma^3)
\delta(x-y)\}.
\end{eqnarray}
As demonstrated in Ref.\cite{invariant}, the $\sigma^3$ part
vanishes automatically for fermion loops with four or less
external vertices, and hence does not contribute to vacuum
polarization diagrams.

With respect to the Euclidean\footnote{In continuing to Euclidean
signature $(+,+,+,+)$ our Dirac matrices satisfy
${\gamma^\mu}^\dagger = \gamma^\mu$.} inner product $\langle X|Y
\rangle = \int d^4x \, e X^\dagger Y$, the full curved space Dirac
operator obeys
$(i{D\kern-0.15em\raise0.17ex\llap{/}\kern0.15em\relax})^\dagger
= i{D\kern-0.15em\raise0.17ex\llap{/}\kern0.15em\relax} +
2i{B\kern-0.15em\raise0.17ex\llap{/}\kern0.15em\relax}$, and the
positive-definite operator which appears in the regulator
function $f$ is
$i{D\kern-0.15em\raise0.17ex\llap{/}\kern0.15em\relax}
(i{D\kern-0.15em\raise0.17ex\llap{/}\kern0.15em\relax})^\dagger$.
In computing the vacuum polarization $\Pi^{\mu\nu}$ defined as
\begin{equation}
\int {{d^4k}\over{(2\pi)^4}}\, \Pi^{\mu\nu}{e^{ik.(x-y)}} \equiv
\left.{{\delta \langle J^{\mu}_L(x)\rangle}\over{\delta
C_{\nu}(y)}} \right|_{C_\alpha =0},
\end{equation}
we need retain only terms linear in $\tilde A_\mu$ and $B_\mu$ in
the regularized current. So displaying just the relevant terms,
\begin{eqnarray}
\langle J^{\mu}_L(x)\rangle_{\rm Reg} =&& \lim_{x \rightarrow
y}{\rm Tr}\{\frac{1}{2}\gamma^\mu\frac{1}{2}(1-\gamma^5) \sum_n
(-1)^n [... +{{m_n}\over{(-\Box + m^2_n)}}
(i{B\kern-0.15em\raise0.17ex\llap{/}\kern0.15em\relax}
+{{A}\kern-0.15em\raise0.17ex\llap{/}\kern0.15em\relax}
\gamma^5){m_n\over{(-\Box + m^2_n)}} \cr
\nonumber\\
&+&{{i{\partial\kern-0.15em\raise0.17ex\llap{/}\kern0.15em\relax}}
\over{(-\Box +
m^2_n)}}(i{B\kern-0.15em\raise0.17ex\llap{/}\kern0.15em\relax} -
{{A}\kern-0.15em\raise0.17ex\llap{/}\kern0.15em\relax}
\gamma^5){{i{\partial\kern-0.15em\raise0.17ex\llap{/}\kern0.15em\relax}}
\over{(-\Box + m^2_n)}} +...]\delta(x-y)\}
\end{eqnarray}
where $\Box =\partial_\mu \partial^\mu$. After moving the
$\gamma^5$ associated with ${\tilde A}_\mu$ to the left, this
works out to be
\begin{eqnarray}
\langle J^{\mu }_L(x)\rangle_{\rm Reg} =&& \lim_{x \rightarrow
y}{\rm Tr}\{\frac{1}{2}\gamma^\mu \frac{1}{2}(1-\gamma^5)\sum_n
(-1)^n [ ... + {{m_n}\over{(-\Box + m^2_n)}}
{C\kern-0.15em\raise0.17ex\llap{/}\kern0.15em\relax}
{m_n\over{(-\Box + m^2_n)}}\cr
\nonumber\\
&&+{{i{\partial\kern-0.15em\raise0.17ex\llap{/}\kern0.15em\relax}}
\over{(-\Box +
m^2_n)}}{C\kern-0.15em\raise0.17ex\llap{/}\kern0.15em\relax}
{{i{\partial\kern-0.15em\raise0.17ex\llap{/}\kern0.15em\relax}}
\over{(-\Box + m^2_n)}} +...]\delta(x-y)\},
\end{eqnarray}
${C\kern-0.15em\raise0.17ex\llap{/}\kern0.15em\relax} =
i{B\kern-0.15em\raise0.17ex\llap{/}\kern0.15em\relax} +
{{A}\kern-0.15em\raise0.17ex\llap{/}\kern0.15em\relax}$ is the
only torsion combination that appears in the final result. It is
crucial to observe the two displayed terms within square brackets
have the same sign. In the result for internal gauge currents a
sign difference occurs (see Ref. \cite{masstor} for the explicit
comparisons). This difference with the polarization of internal
gauge fields and torsion can be traced precisely to the axial
vector ${\tilde A}_\mu$ and the non-hermitian $iB_\mu$ couplings
to fermions. Thus the origin of non-transverse torsion
polarization for the Weyl theory is essentially the same as that
for the bispinor theory discussed earlier. The relevant Feynman
diagram processes for the polarization can also be read off from
the above expression of the current. The first term arises from
nontrivial ${\overline\Psi}_L{\overline\Psi}_L$ and
${\Psi_L}{\Psi_L}$ propagators due to Majorana regulator masses,
while the second is associated with $\Psi_L{\overline\Psi}_L$
propagators. Full expressions of regularized Weyl propagators can
be found in Ref.\cite{invariant}. Following detailed computations
in Ref. \cite{masstor}, the polarization is
\begin{equation}
\Pi^{\mu\nu} = (A^{\mu\nu}_T + A^{\mu\nu}_L)d,
\end{equation}
where the transverse piece
\begin{equation}
A^{\mu\nu}_T ={1\over {24\pi^2}} (k^\mu k^\nu -
g^{\mu\nu}k^2)[\ln ({{k^2}\over{\Lambda^2}}) - \frac{5}{3} + 2
\ln ({\pi \over 2})]
\end{equation}
is, apart from a $Tr\{T_aT_b\}$ internal gauge factor for
$\Pi^{\mu\nu}_{ab}$, identical to the result obtained for
internal gauge couplings\cite{Aoki}. The nontrivial anomalous
longitudinal component is
\begin{equation}
A^{\mu\nu}_L = -g^{\mu\nu}[{{7\zeta(3)} \over {4\pi^4}}\Lambda^2
+ k^2 \sum_{n=0}{{(2^{2n-1} -1)n! B_{2n}
\pi^{2n-2}{\sqrt\pi}}\over {2^{2n+4}(2n)!\Gamma(n +
\frac{5}{2})}}\left({k^2 \over \Lambda^2}\right)^n]
\end{equation}
written in terms of Bernoulli numbers, gamma and zeta functions and a convergent power series in
$(k^2/\Lambda^2)$.

In the effective propagator with vacuum polarization insertions,
it is known that a non-trivial longitudinal polarization causes a
shift to a physically massive pole, even if the bare propagator
is massless in the beginning (see, for instance,
Ref.\cite{PeskinBook}). Since $\Pi^{\mu\nu}$ is the Fourier
transform of $\left.{{\delta^2 \Gamma_{eff.}} \over{{\delta
C_\nu}{\delta C_\mu}}}\right|_{C=0}$, it follows that in addition
to the more familiar curvature squared counter term $g^{\mu
\alpha}g^{\nu\beta}(\partial_\mu C_\nu -\partial_\nu C_\mu)
(\partial_\alpha C_\beta -\partial_\beta C_\alpha)$ required by
the logarithmic divergence of the transverse part of
$\Pi^{\mu\nu}$, a counter term proportional to $g^{\mu\nu}C_\mu
C_\nu$ for the longitudinal component of the polarization is also
needed in the Lagrangian when the $\Lambda \rightarrow \infty$
limit is taken. The presence of these terms in the effective
action implies that as a result $C_\mu$ becomes massive and obeys
the Proca equation.

\section{Concluding remarks}

The Weyl action of Eq.\ (\ref{Laction}) would be gauge invariant
under {\it local} $\gamma^5$ and scaling transformations
\begin{equation}
\Psi_L \rightarrow \exp(i\alpha(x)\gamma^5
-\frac{3}{2}\beta(x))\Psi_L = T \Psi_L , \qquad
e{\overline\Psi}_L\gamma^\mu \rightarrow
e{\overline\Psi}_L\gamma^\mu T^{-1} ,
\end{equation}
with $T(x)=\exp[-(i\alpha +\frac{3}{2}\beta)]$ if we {\it
pretend} that $C_\mu = (iB_\mu + A_\mu)$ is a complex Abelian
gauge connection which transforms as
\begin{equation}
C_\mu \rightarrow T C_\mu T^{-1} - T i\partial_\mu T^{-1}  \quad
{\rm i.e.} \quad A_\mu \rightarrow A_\mu +\partial_\mu \alpha
,\qquad B_\mu \rightarrow B_\mu - \frac{3}{2}\partial_\mu\beta.
\end{equation}
Note that $B_\mu$ comes with an $i$ in the complex combination $C_\mu$ because, unlike $\gamma^5$ rotations, the
group parametrized by $\exp(-\frac{3}{2}\beta)$ is noncompact rather than $U(1)$. However massive regulators
break both of these symmetries, and the current $J^\mu_L$ coupled to $C_\mu$ is not conserved. The full Weyl
theory however exhibits no inconsistencies because these invariances are really not gauged as local symmetries.
The theory is on the other hand diffeomorphism, and local internal gauge and Lorentz invariant, with internal
symmetries gauged by $W_{\mu a}$ and local Lorentz invariance by the spin connection $A_{\mu AB}$. $C_\mu$ is in
fact a composite which transforms covariantly under general coordinate transformations, and is invariant under
local Lorentz and gauge transformations.

The appearance of a mass term for $C^\mu$ does have an important
consequence. The combination $C_\mu$ is complex, therefore the
counter terms of the Lorentzian signature Lagrangian required by
vacuum polarization diagrams include anti-hermitian cross terms
which are Lorentz invariant, but CPT-odd. In a related work
\cite{cpt} it was pointed out that a truly Weyl theory which
preserves the chirality of the gravitational interaction should
violate CPT through $B_\mu$ effects; and that these signatures of
chiral gravity will already be manifest at the level of quantum
field theory in background curved spacetimes. An example is
precisely the vacuum polarization diagram with two complex
left-handed spin connection vertices. The flat vierbein limit
with nontrivial $C_\mu$ evaluated here indeed confirms the
presence of these CPT-violating terms in the effective action.

In general, we may decompose the torsion as $T_{\mu\nu A} = {2
\over 3}(B_\mu e_{\nu A} - B_\nu e_{\mu A})
             + {1\over 3}\epsilon_{\mu\nu A\alpha}\tilde{A^\alpha}
             + Q_{\mu\nu A}$,
where $Q_{\mu\nu A} $ is constrained by $e^{\nu A} Q_{\mu\nu A} =
\epsilon^{\mu\nu A \alpha} Q_{\mu\nu A}=0.$ Since $Q_{\mu\nu A}$
does not couple to fermions, the spin connection in the fermion
coupling can be restricted to
\begin{equation}
A_{\mu AB} = \omega_{\mu AB} + {2 \over 3}(B_\nu e^\nu _A e_{\mu
B} - B_\nu e^\nu_B e_{\mu A}) - {1\over
6}{\epsilon}_{AB\mu\alpha}\tilde{A^\alpha}
\end{equation}
As a result, the Samuel-Jacobson-Smolin action\cite{Ash} (which is the (anti)-self-dual part of the
Einstein-Hilbert-Palatini action), ${i\over{8\pi G}}\int e^A \wedge e^B \wedge {1\over 2}(i*+1)F_{AB}$, is
equivalent to ${1\over {16\pi G}}\int d^4x (eR - 4i\partial_\mu (eC^\mu) +{8\over 3} C_\mu C^\mu)$. So modulo the
total divergence term which is not reproduced by perturbation theory, the $C_\mu C^\mu$ counter term required by
the vacuum polarization is the same as the Samuel-Jacobson-Smolin action in the teleparallel limit. We can
therefore expect the relevant mass term for $C_\mu$ in the renormalized effective action of Weyl fermions coupled
to full-fledged background gravitational fields to be the corresponding quadratic term of the full
Samuel-Jacobson-Smolin action; and the coefficient of $C_\mu C^\mu$ in the effective action is thus ${1 \over
{(6\pi G_{renor.})}}$, implying a natural Planck-scale mass.

Phenomenological consequences of massive axial torsion modes have been discussed before\cite{Shap}. In our
present context, we wish to pursue the extent to which chirality can be used as a defining characteristic of
particle interactions, including gravity.  That context required us to use Weyl spinors, and also the Ashtekar
formulation of gravity. A net consequence is that both axial torsion as well as vector torsion trace are needed,
but with a relative phase which ruins CPT invariance because of the chiral nature of the fields. Massive modes
appear as consequences of the anomalous non-conservation of the current to which these torsion modes are coupled.
At low energies compared to the torsion mass, the fermion-torsion interaction reduces to a four-fermion coupling.
Present high energy experimental data on four-fermion vertices sets the lower bound for torsion mass at above the
200GeV scale\cite{Shap}.

The question of mass generation via anomalies has had a storied
past. In the Schwinger model in 2D, the physical degree of
freedom of the photon is equivalent to a free massive
boson\cite{Schwinger,Brown} since the interaction term can be
transformed away by an axial rotation. The mass of the boson
stems from the ABJ anomaly, which gives rise to an infra-red pole
in the polarization tensor. On the other hand, the chiral
Schwinger model in 2D is anomalous albeit exactly solvable. The
resultant photon mass, while still finite, carries an ambiguity
as the previous condition of gauge invariance is now absent. It
can be made to vanish, while preserving the (V-A) form for the
coupling, but we then lose unitarity \cite{Jackiw}.
This paper discusses the corresponding chiral situation in 4D,
but without loss of any local gauge invariance. By retaining
explicitly all local gauge symmetries and the holomorphic
dependence on the left-handed spin connection in the
regularization, we end up with a vacuum polarization tensor that
is non-transverse, and gives a mass to $C_\mu =iB_\mu + A_\mu$.
These complex torsion modes are massive because of ABJ and scaling
anomalies, with generated masses of the order of the regulator
scale. Since these are the only modes that can couple to spin
$\frac{1}{2}$ fermions, large torsion masses, or high cut-off
scales in the context of effective field theories, naturally
explain why torsion and its associated effects, including CPT
violations from $B_\mu$ couplings, have so far escaped detection.

\acknowledgments

The research for this work has been supported in part by funds from the U.S. Department of Energy under Grant No.
DE-FG05-92ER40709, and the National Science Council of Taiwan under Grant Nos. NSC 90-2112-M-006-012 and NSC
91-2112-M-006-018.


\end{document}